\documentclass[12pt]{article}
\usepackage{amsmath}
\usepackage{amssymb}
\usepackage{graphicx}

\input{epsf}

\begin{document}

\begin{center}
{\Large \bf  Spin network setting of topological\\ quantum computation}
\end{center}
\vspace{6pt}

\begin{center}
{\large
{\sl Annalisa Marzuoli}}\\

\vspace{4pt}
{\small{Dipartimento di Fisica Nucleare e Teorica,
Universit\`a degli Studi di Pavia and 
Istituto Nazionale di Fisica Nucleare, Sezione di Pavia, \\
via A. Bassi 6, 27100 Pavia (Italy) \\
annalisa.marzuoli@pv.infn.it}} \\
\end{center}

\begin{center}
{\large
{\sl Mario Rasetti}}\\

\vspace{4pt}
{\small{Dipartimento di Fisica and Istituto Nazionale di Fisica della Materia,\\
Politecnico di Torino,\\
corso Duca degli Abruzzi 24, 10129 Torino (Italy) \\
mario.rasetti@polito.it}} \\
\end{center}

\vspace{6pt}

\begin{center}
{\bf Abstract}
\end{center}

{\small{The spin network simulator model represents a bridge between
(generalised) circuit schemes for standard quantum computation
and approaches based on notions from Topological 
Quantum Field Theories (TQFTs). The key tool is provided by the fiber space structure
underlying the model which exhibits combinatorial properties closely
related to $SU(2)$ state sum models, widely employed in discretizing TQFTs 
and quantum gravity in low spacetime dimensions.}}

\vspace{6pt}

\noindent PACS numbers: 03.67.Lx, 03.65.Fd, 11.10.Kk \\
{\em Keywords}: Quantum Computation, Spin Networks, Topological 
Quantum Field Theory.

\vspace{12pt}

We discuss here some aspects of a novel setting for quantum computation \cite{MaRa1} \cite{MaRa2}
based on the (re)coupling theory of $SU(2)$ angular momenta \cite{BiLo9} \cite{Russi}. 
On the one hand, the $''$spin network simulator$''$ can be thought of as 
a non--Boolean generalisation of the 
{\em quantum circuit} model, with unitary gates expressed in terms of:\\
{\bf i)} recoupling coefficients ($3nj$ symbols) between inequivalent binary
coupling schemes of $N\equiv(n+1)$ $SU(2)$--angular momenta ($j$--gates);\\
{\bf ii)} Wigner rotations in the eigenspace of the total angular momentum ($M$--gates).
The picture does  contain the Boolean case  as the particular case when 
all $N$ angular momenta are spin $\frac{1}{2}$.
The combinatorial structure of the model, on the other hand, closely resembles $SU(2)$
$''$state sum models$''$ employed in discretized approaches to 2 and 3--dimensional 
Topological Quantum Field Theories \cite{Kau} (TQFT) and Quantum Gravity \cite{ReWi}. 
An explicit mapping relating the spin network to the topological approach to quantum
computation based on modular functors \cite{FrLaWa} can be established as illustrated 
in Ref. 2, Section 6.

The basic ingredients of the kinematics of the simulator are 
computational Hilbert spaces (to be defined below) 
equipped with all unitary gates of types {\bf i)} and {\bf ii)}
we may consistently apply to transform vectors 
belonging to such spaces. A specific computation will be implemented by picking up
an $''$input$''$ state to get an $''$output$''$  generated by acting 
with a particular (finite) sequence of elementary gates. 
An elementary operation will be carried out in one unit of
time since it is assumed that the intrinsic time variable of the simulator has a 
discrete character.\\
The architecture of the spin network is modelled as an $SU(2)$ fiber space structure
over a discrete base space $V$
\begin{equation}\label{1}
(V,\,\mathbb{C}^{2J+1},\,SU(2)^J)_n
\end{equation}
which encodes all possible computational Hilbert spaces as well
as  gates for any fixed number $N=n+1$ of incoming angular momenta.

The base space $V\;\doteq\;\{v(\mathfrak{b})\}$ represents the vertex set of a regular,
$2(n-1)$--valent graph $\mathfrak{G}_n(V, E)$ with cardinality $|V| = (2n-1)!!$.
There exists a 1:1 correspondence
\begin{equation}\label{2}
\{v(\mathfrak{b})\}  \longleftrightarrow \{\mathcal{H}^J_n\,(\mathfrak{b})\}
\end{equation}
between the vertices of $\mathfrak{G}_n(V, E)$ and the computational Hilbert spaces 
of the simulator. For a given value of $n$, $\mathcal{H}^J_n(\mathfrak{b})$ is the simultaneous
eigenspace of the squares of $2n+1$ Hermitean, mutually commuting angular
momentum operators, namely 
$${\bf J}_1,\;{\bf J}_2,\;{\bf J}_3,\ldots,{\bf J}_{n+1}\,\equiv \,\{{\bf J}_i\};\;\;\;\; 
{\bf J}_1\,+\,{\bf J}_2\,+\,{\bf J}_3\,+\ldots+{\bf J}_{n+1}\;\doteq\;{\bf J};$$
\begin{equation}\label{3}
{\bf K}_1,\,{\bf K}_2,\,{\bf K}_3,\,\ldots,\,{\bf K}_{n-1}\,\equiv \,\{{\bf K}_h\}
\end{equation}
and of the operator $J_z$ (the projection of the total angular momentum $\bf{J}$
along the quantization axis). The associated quantum numbers are 
$j_1, j_2,\ldots,j_{n+1}$ $;\,J;$ $ k_1,k_2,\ldots,$ $k_{n-1}$ and $M$, where $-J \leq M
\leq$ in integer steps. If
${\cal H}^{j_1}\otimes$ ${\cal H}^{j_2}\otimes\cdots$ $\otimes 
{\cal H}^{j_{n}}\otimes {\cal H}^{j_{n+1}}$
denotes the factorized Hilbert space, namely the $(n+1)$--fold tensor product 
of the individual eigenspaces of the $({\bf J}_i)^2\,$'s, the operators 
${\bf K}_h$'s represent intermediate angular momenta generated, through Clebsch--Gordan series, 
whenever a pair of ${\bf J}_i$'s are (binarily) coupled. As an example, by coupling
sequentially the ${\bf J}_i$'s according to the scheme
$(\cdots(({\bf J}_1+{\bf J}_2)+{\bf J}_3)+\cdots+{\bf J}_{n+1})$ $={\bf J}$ -- which generates
$({\bf J}_1+{\bf J}_2)={\bf K}_1$,
$({\bf K}_1+{\bf J}_3)={\bf K}_2$, and so on --
we should get a binary bracketing structure of the type
$(\cdots((({\cal H}^{j_1}\otimes{\cal H}^{j_2})_{k_1}$ $\otimes{\cal H}^{j_3})_{k_2}
\otimes$ $\cdots \otimes
{\cal H}^{j_{n+1}})_{k_{n-1}})_J$, where we add an overall  bracket labeled by the quantum
number of the total angular momentum $J$. Note that, as far as $j_i$'s
 quantum numbers are involved, any value belonging to 
 $\{0,1/2,1,3/2,\ldots \}$ is allowed, while the ranges of the $k_h$'s are suitably 
 constrained by Clebsch--Gordan decompositions
 ({\em e.g.} if $({\bf J}_1+{\bf J}_2)={\bf K}_1$ $\Rightarrow$ $|j_1-j_2| \leq$
 $k_1 \leq j_1+j_2$).
We denote a binary coupled basis of $(n+1)$ angular
momenta in the $JM$--representation 
and the corresponding Hilbert space appearing in (\ref{2}) according to
$$\{\,|\,[j_1,\,j_2,\,j_3,\ldots,j_{n+1}]^{\mathfrak{b}}\, ;k_1^{\mathfrak{b}\,},\,k_2^{\mathfrak{b}\,}
,\ldots,k_{n-1}^{\mathfrak{b}}\, ;\,JM\, \rangle,\;
-J\leq M\leq J \}$$
\begin{equation}\label{4}
=\;{\cal H}^{J}_{\,n}\;(\mathfrak{b})\;\doteq\;\mbox{span}\;\{\;|\,\mathfrak{b}\,;JM\,\rangle_n\,\}\;,
\end{equation}
where  the string inside $[j_1,\,j_2,\,j_3,\ldots,j_{n+1}]^{\mathfrak{b}\,}$ is not necessarily
an ordered one, $\mathfrak{b}$ indicates the current binary bracketing structure and 
the $k_h$'s are uniquely associated with the chain of pairwise couplings selected by $\mathfrak{b}$.
For a given value of $J$
each $\mathcal{H}^J_n (\mathfrak{b})$ has dimension $(2J + 1)$ over 
$\mathbb{C}$ and thus there exists one isomorphism
\begin{equation}\label{5}
\mathcal{H}^J_n (\mathfrak{b})\;\;\; \cong _{\,\mathfrak{b}}\;\;\; \mathbb{C}^{2J+1}
\end{equation}
for each admissible binary coupling scheme $\mathfrak{b}$ of $(n + 1)$ incoming spins.
The vector space $\mathbb{C}^{2J+1}$ is interpreted as the typical fiber attached to each vertex
$v(\mathfrak{b}) \in V$ of the fiber space structure (\ref{1}) through the isomorphism (\ref{5}).

For what concerns unitary operations acting on the computational
Hilbert spaces (\ref{4}), we examine first $j$--gates associated with recoupling 
coefficients ($3nj$ symbols) of $SU(2)$ as anticipated in
{\bf i)}. It can be shown \cite{BiLo9} that any such coefficient can be 
splitted into $''$elementary$''$ $j$--gates, the Racah transforms, possibly apart from
phases and weight factors. A Racah transform applied to a basis vector is defined formally as
\begin{equation}\label{6}
{\cal R}\;:\,| \dots (\,( a\,b)_d \,c)_f \dots;JM \rangle\; \mapsto \;
\,|\dots( a\,(b\,c)_e\,)_f \dots;JM \rangle, 
\end{equation}
where we are using here Latin letters $a,b,c,\ldots$ to denote both incoming ($j_i\,$'s
in the previous notation) 
and intermediate ($k_h\,$'s) spin quantum numbers.
The explicit expression of (\ref{6}) reads
$$|(a\,(b\,c)_e\,)_f\,;M\rangle$$
\begin{equation}\label{7} 
=\sum_{d}\,(-1)^{a+b+c+f}\; [(2d+1)
(2e+1)]^{1/2}
\left\{ \begin{array}{ccc}
a & b & d\\
c & f & e
\end{array}\right\}\;|(\,(a\,b)_d \,c)_f \,;M\rangle,
\end{equation}
where there appears the Racah--Wigner $6j$ symbol of $SU(2)$ and $f$ plays the role
of the total angular momentum quantum number.
Recall that the square of the $6j$ symbol in (\ref{7}) represents the 
probability that a system prepared in the state $|(\,(a\,b)_d \,c)_f \,;M\rangle$
will be measured in the state $|(a\,(b\,c)_e\,)_f\,;M\rangle$.
Moreover, Racah transforms are the key ingredients to complete
the construction of the Rotation graph $\mathfrak{G}_n(V, E)$ introduced 
above (note that in this case  $''$Rotation$''$
refers to a topological operation on binary tree structures and not to rotation
operators referred to in {\bf ii)}, see Ref. 2, Appendix A).
The edge set $E = \{e\}$ of $\mathfrak{G}_n(V, E)$ is a subset of the Cartesian
product $(V \times V )$ selected by the action of elementary $j$--gates. More precisely, 
an (undirected) arc between two vertices $v(\mathfrak{b})$ and $v(\mathfrak{b}')$
\begin{equation}\label{8}
e\,(\mathfrak{b},\mathfrak{b}')\;\doteq \;(v(\mathfrak{b}),\, v(\mathfrak{b}')) 
\;\in \;(V \times V)
\end{equation}
exists if, and only if, the underlying Hilbert spaces are related to each other by 
an elementary unitary operation of the type (\ref{6}). 
Note also that elements in $E$ can be considered as mappings
$$
(V\,\times\,\mathbb{C}^{2J+1})_n\; \longrightarrow\,
(V\,\times\,\mathbb{C}^{2J+1})_n
$$
\begin{equation}\label{9}
\;\;\;\;\;\;\;(v(\mathfrak{b}),\,\mathcal{H}^J_n (\mathfrak{b})\,)\, \mapsto\,
(v(\mathfrak{b}'),\,\mathcal{H}^J_n (\mathfrak{b}')\,)
\end{equation}
connecting each given decorated vertex to one of its nearest $2(n-1)$ 
vertices and thus define a $''$transport 
prescription in the horizontal sections$''$ belonging to the total space
$(V \times \mathbb{C}^{2J+1})_n$ of the fiber bundle (\ref{1}). 
The crucial feature that characterises the graph $\mathfrak{G}_n(V, E)$ arises from 
compatibility conditions satisfied by $6j$ symbols appearing in (\ref{7}).
The Racah (triangular) identity, the Biedenharn--Elliott (pentagon) identity
and the orthogonality conditions for $6j$ symbols \cite{Russi}
ensure that any simple path in $\mathfrak{G}_n(V, E)$ with fixed endpoints can 
be freely deformed into any other, providing identical quantum transition amplitudes
(and of course probabilities)
{\em at the kinematical level} ({\em cfr.} Ref. 2, Section 3.1 for more details).

To complete the description of the structure 
$(V,\,\mathbb{C}^{2J+1},\,SU(2)^J)_n$ we call into play $M$--gates which act
on the angular dependence of vectors in $\mathcal{H}^J_n (\mathfrak{b})$.
By expliciting such dependence according to
\begin{equation}\label{10}
\mathcal{H}^J_n (\mathfrak{b})\;\doteq\;\mbox{span}\,
 \{\;|\mathfrak{b};\theta,\, \phi;\,
JM\rangle_n\,\},
\end{equation}
we write the action of a rotation on a basis vector as
\begin{equation}\label{11}
|\mathfrak{b};\theta',\, \phi';\,M'J\,\rangle_n\;=\;
\sum_{M=-J}^{J}\;D^J_{MM'}\,(\alpha \beta \gamma)\,
|\mathfrak{b};\theta,\, \phi;\,JM\,\rangle_n\;,
\end{equation}
where $(\theta, \phi)$ and $(\theta', \phi')$ are polar angles in the original and rotated coordinate 
systems, respectively.
$D^J_{MM'}\,(\alpha \beta \gamma)$
are Wigner rotation matrices in the $JM$ representation expressed in terms of Euler angles 
$(\alpha \beta \gamma)$ which form a group under composition \cite{Russi} (see also Ref. 2, 
Section 3.2 and Appendix B1, for a discussion on the possibility of selecting $''$elementary$''$
$M$--gates in some specific cases). 
In what follows we shall assume for simplicity that, for given $n$ and $J$,
it takes just one unit of time to perform one rotation. The shorthand notation
$SU(2)^J$ employed in (\ref{1}) actually refers to the group of W--rotations,
which in turn can be interpreted as the automorphism group
of the $''$fiber$''$ $\mathbb{C}^{2J+1}$. Since rotations in the $JM$ representation do
not alter the binary bracketing structure of vectors in computational Hilbert spaces
we can interpret actions of W--matrices as $''$transport prescriptions
along the fiber$\,''$.

The framework outlined above makes it manifest the fact that we can
switch {\em independently} 
$j$ and $M$--gates without mixing spin and magnetic quantum numbers.
This feature, which relies on the discreteness of the base space $V$  and on the $''$triviality$\,''$
of the total space $\,(V\times\mathbb{C}^{2J+1})_n\,$, makes it easy with kinematics,
although the simulator will exhibit complex behaviors at the dynamical level.
 
A pictorial representation 
of the computational fiber space 
$(V,\,\mathbb{C}^{2J+1},$ $\,SU(2)^J)_n$ for $(n+1)=4$ incoming spins is sketched in Fig. 1.
Each vertex is associated, through the mapping (\ref{2}), with a binary coupled Hilbert space
$\mathcal{H}^J_n (\mathfrak{b})$ 
depicted as a plane. Edges of the graph represent Racah transforms (\ref{6})
which actually move each space viewed as a whole into one of its 
nearest $2(n-1)=4$ ones.
Inside each $\mathcal{H}^J_n (\mathfrak{b})$ $\cong \mathbb{C}^{2J+1}$ (see (\ref{5}))
we can pick up a particular vector (as shown in the drawing) and rotate
its components by means of the Hermitean conjugate of the matrix 
$D^J_{MM'}\,(\alpha \beta \gamma)$ introduced in (\ref{11}) for some choice of
$(\alpha \beta \gamma)$.

\vfill
\eject

\begin{figure}
\begin{center}
\includegraphics[bb= 0 250 600 650, scale=0.5]{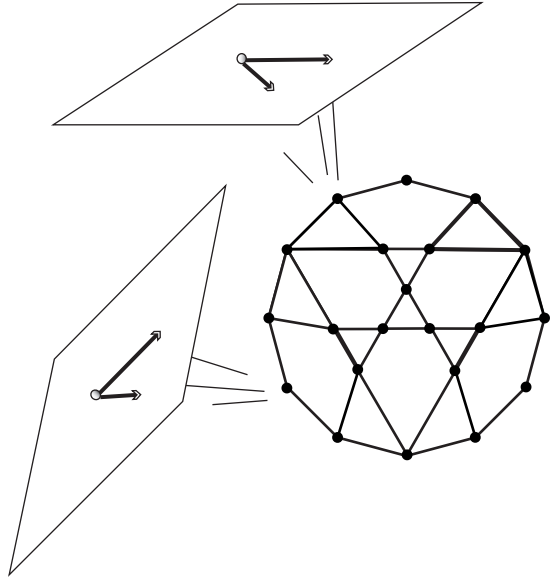}
\end{center}
\caption{The fiber space structure of the spin
network simulator for $(n+1)=4$ incoming spins.
Vertices and edges on the perimeter of the
Rotation graph $\mathfrak{G}_3 (V,E)$ have to be identified through the antipodal map.
A couple of vertices are $''$blown up$''$ into planes representing the associated 
computational Hilbert spaces.} 
\label{figura1}
\end{figure}

The dynamical behavior of the simulator in working as a quantum circuit can be discussed in terms
of {\em directed paths} in the fiber space structure 
$(V,\,\mathbb{C}^{2J+1},\,SU(2)^J)_n$. 
By a directed path $\mathcal{P}$
we mean a (time) ordered sequence 
\begin{equation}\label{12}
|\mathfrak{v}_{\mbox{in}}\,\rangle_n\equiv
|\mathfrak{v}_{0}\,\rangle_n\rightarrow
|\mathfrak{v}_{1}\,\rangle_n\rightarrow\cdots\rightarrow
|\mathfrak{v}_{s}\,\rangle_n\rightarrow\cdots\rightarrow
|\mathfrak{v}_{L}\,\rangle_n\equiv
|\mathfrak{v}_{\mbox{out}}\,\rangle_n\;,
\end{equation}
where we use the shorthand notation $|\mathfrak{v}_{s}\rangle_n$ for computational states and
$s=0,1,2,\ldots ,L$ is the lexicographical labelling of the states along the 
given path with fixed endpoints. 
$L$ is the length of the path $\mathcal{P}$, which turns out to be proportional to the 
time duration of the computation process $L \cdot \tau \doteq T$ in terms of the 
discrete time unit $\tau$. The integer $L$ characterising the particular directed path 
in (\ref{12}) equals the
 number of time--ordered elementary operations (computational steps) needed to 
get $|\mathfrak{v}_{\mbox{out}}\,\rangle_n$ from
$|\mathfrak{v}_{\mbox{in}}\,\rangle_n$ following the path $\mathcal{P}$. An elementary 
computational
step, represented by an arrow in (\ref{12}), is either a Racah transform or a W--rotation.
A circuit--type computation consists in evaluating the expectation value of the unitary
operator $\mathfrak{U}_{\mathcal{P}}$
associated with the path $\mathcal{P}$, namely
\begin{equation}\label{13}
\langle \mathfrak{v}_{\mbox{out}}\,|\,\mathfrak{U}_{\mathcal{P}}\,|\,
\mathfrak{v}_{\mbox{in}}\,\rangle_n.
\end{equation}
By taking advantage of the possibility of decomposing 
$\mathfrak{U}_{\mathcal{P}}$
 uniquely into an ordered sequence of elementary gates, (\ref{13}) becomes
\begin{equation}\label{14}
\langle \mathfrak{v}_{\mbox{out}}\,|\,\mathfrak{U}_{\mathcal{P}}\,|\,
\mathfrak{v}_{\mbox{in}}\,\rangle_n\;=\;
\lfloor\,
\prod_{s=0}^{L-1}\,
\langle \mathfrak{v}_{s+1}\,|\,\mathcal{U}_{s,s+1}\,|\,
\mathfrak{v}_{s}\,\rangle_n\;\rfloor_{\mathcal{P}}
\end{equation}
with $L\equiv L(\mathcal{P})$ for short. The symbol  
$\lfloor \; \rfloor_{\mathcal{P}}$ denotes the ordered 
product along the path $\mathcal{P}$ 
and each elementary operation 
is now denoted by $\mathcal{U}_{s, s+1}$ $(s =0,1,2, \ldots L(\mathcal{P}))$
to stress its $''$one--step$\, ''$ 
character with respect to computation.
 Consequently, each elementary transfer matrix in (\ref{14}) 
turns out to be associated with a
 local Hamiltonian operator arising from
\begin{equation}\label{15}
\langle \mathfrak{v}_{s+1}\,|\,\mathcal{U}_{s,s+1}\,|\,
\mathfrak{v}_{s}\,\rangle_n\;=\;\exp\,\{i\, \mathbf{H}_n\,(s,s+1)\cdot \tau\}
\end{equation}
and representing the unitary evolution of the simulator in one unit of its 
intrinsic time variable $\tau$. We indicate with the 
shorthand notation $(s,s+1)$ the dependence of $\mathbf{H}_n$ on its variables to make clear 
the local nature of this operator with respect to the computational space 
$(V,\,\mathbb{C}^{2J+1},\,SU(2)^J)_n$.
When (\ref{15}) is inserted in (\ref{14}), such {\em virtual} 
Hamiltonians generally do not commute with each 
other but nonetheless the whole computational process may be identified with a well defined 
unitary evolution of the simulator, driven by a poly--local Hamiltonian,
in the internal time interval $T=L(\mathcal{P}) \cdot \tau$.

Different types of evolutions 
can be grouped into $''$computing classes$\, ''$ based on the choice of 
gates that a particular program ({\em i.e.} a collection of directed paths) has to employ. 
A computing class that alternates $j$ and $M$--gates would be the
most general one (as explained in Ref. 2, Sect. 4.2). However, two particular classes
to be illustrated below come out to be quite interesting 
in view of the possibility of relating them to other models
for quantum computation. 

An $M$--{\em computing class} contains programs which employ only (finite sequences of)
$M$--gates 
in their associated directed paths and it is 
not difficult to realize that such kind of computation, when applied to $N$ 
$\frac{1}{2}$--spins, reproduces the usual Boolean quantum circuit. 

A $j$--{\em computing class} includes programs which employ only $j$-gates at
 each computational step. This class is particularly interesting since it shares 
many features with suitable types of discretized TQFTs, the so--called
state sum models\cite{Kau}\cite{ReWi}\cite{CaMa},
 as we already noticed in Ref. 1 and developed in details in Ref. 2, Sect. 5.
 Now the combinatorial structure of Rotation graphs becomes prominent owing to the existence
 of a 1:1 correspondence between allowed elementary operations and the edge set $E$ of 
the graph $\mathfrak{G}_n(V,E)$, for any $n$. We denote the unitary operator associated with
a program $\mathcal{P}$ in this class by
\begin{equation}\label{16}
\mathcal{U}_{3nj}\, :\,
|\mathfrak{v}_{\mbox{in}}\rangle_n\;\longrightarrow\;|\mathfrak{v}_{\mbox{out}}\rangle_n\;, 
\end{equation}
where, in a circuit--type setting, $|\mathfrak{v}_{\mbox{in}}\rangle_n$
is fixed and $|\mathfrak{v}_{\mbox{out}}\rangle_n$ is an accepted state. 
However, in the $j$--computing class one
may address other types of problems. For instance, once selected two states, say
$|\mathfrak{v}_{\mbox{in}}\rangle_n$ and
$|\mathfrak{v}_{\mbox{out}}\rangle_n$, 
we may consider all possible paths $\mathcal{P}$ that compute
$|\mathfrak{b}_{\mbox{out}}\rangle_n$
as the result of the application of some
$\mathcal{U}_{3nj}$
to $|\mathfrak{b}_{\mbox{in}}\rangle_n$. 
The functional which takes care 
of such multiple choices is a $''$path sum$\, ''$ (a discretized Feynman's path integral) 
which may be written formally as
\begin{equation}\label{17}
\mathbf{Z}[\mathfrak{v}_{\mbox{in}},\,\mathfrak{v}_{\mbox{out}}]\;=\;
\sum_{\mathcal{P}}\,
W_{\mathcal{P}}\,
\langle\mathfrak{v}_{\mbox{out}}\,|\,
\mathcal{U}_{3nj}\,|\,
\mathfrak{v}_{\mbox{in}}\,\rangle_n\,,
\end{equation}
\noindent where the summation is over all paths with fixed endpoints and $W_{\mathcal{P}}$ is a 
weight to be assigned to each particular path. The explicit form of such functional
would contain sums over intermediate spin variables $k_h\,$'s of products of $6j$ symbols
with suitable weights (dimensions of $SU(2)$ irreps labeled by
$k_h\,$'s) and phase factors.
Note that if we give the same 
weight, say $W_{\mathcal{P}} =1$, to each path, then the results on 
equi--probable amplitudes derived from the algebraic identities for $6j$ symbols
ensure that the functional (\ref{17})
is a combinatorial invariant, namely it is actually independent of the particular 
path connecting 
$\mathfrak{v}_{\mbox{in}}$ and $\mathfrak{v}_{\mbox{out}}$. 
The situation in this case resembles what happens in studying classes of combinatorial 
invariants  ($SU(2)$--state sum functionals) for triangulated manifold
originally defined in dimension 3 (see Refs. 5,6 for reviews)
and extended (in all dimensions) to manifolds with boundary \cite{CaMa}.
Functionals like (\ref{17}) can be  actually
associated with particular classes of extended
geometrical objects, namely triangulated surfaces and possibly triangulated
3--manifolds. On the other hand, the number of incoming spins must be kept fixed
in a quantum circuit context, and thus
we cannot conclude that such a computing machine  is able to simulate TQFTs or gravity where
the path integrals contain sums over all configurations of fields  (up to regularization).
Other intriguing possibilities open when we insert non trivial weights $W_{\mathcal{P}}$ in
(\ref{17}), since we may naturally address questions about most efficient 
algorithms and time complexity.
For instance we could weigh paths with the inverse of their lengths $L({\cal P})$; 
then the minimum--length path (the optimal algorithm) will be dynamically singled out in 
the path sum. As a matter of fact, even such simple example turns out to be highly 
non trivial owing to the combinatorial complexity of ${\mathfrak{G}}_n (V,E)$ \cite{Belgi}, as pointed out
in Ref. 2 (Sect. 4.3 and Appendix A). 

As a final remark we mention the possibility of exploiting the fiber space structure 
$(V,\,\mathbb{C}^{2J+1},\,SU(2)^J)_n$ to provide a natural discrete--time implementation of holonomic
quantum computation \cite{ZaRa}, at least when the Hamiltonians defined in (\ref{15}) 
exhibit the required degeneracy.

\end{document}